\documentclass[aps,prl,twocolumn,showpacs,superscriptaddress]{revtex4}
\usepackage{graphicx}
\begin{document}


\title{Doping-Enhanced Antiferromagnetism in Ca$_{1-x}$La$_x$FeAs$_2$}


\author{Shinji Kawasaki}
\affiliation{Department of Physics, and Research Center of New Functional Materials for Energy Production, Storage and Transport,Okayama University, Okayama 700-8530, Japan}

\author{Tomosuke Mabuchi}
\affiliation{Department of Physics, and Research Center of New Functional Materials for Energy Production, Storage and Transport,Okayama University, Okayama 700-8530, Japan}

\author{Satoki Maeda}
\affiliation{Department of Physics, and Research Center of New Functional Materials for Energy Production, Storage and Transport,Okayama University, Okayama 700-8530, Japan}

\author{Tomoki Adachi}
\affiliation{Department of Physics, and Research Center of New Functional Materials for Energy Production, Storage and Transport,Okayama University, Okayama 700-8530, Japan}

\author{Tasuku Mizukami}
\affiliation{Department of Physics, and Research Center of New Functional Materials for Energy Production, Storage and Transport,Okayama University, Okayama 700-8530, Japan}

\author{Kazutaka Kudo}
\affiliation{Department of Physics, and Research Center of New Functional Materials for Energy Production, Storage and Transport,Okayama University, Okayama 700-8530, Japan}

\author{Minoru Nohara}
\affiliation{Department of Physics, and Research Center of New Functional Materials for Energy Production, Storage and Transport,Okayama University, Okayama 700-8530, Japan}

\author{Guo-qing Zheng}
\affiliation{Department of Physics, and Research Center of New Functional Materials for Energy Production, Storage and Transport,Okayama University, Okayama 700-8530, Japan}

\affiliation{Institute of Physics and Beijing National Laboratory for Condensed Matter Physics, Chinese Academy of Sciences, Beijing 100190, China}

\date{\today}

\begin{abstract}
In iron pnictides, high temperature superconductivity emerges after suppressing antiferromagnetism by doping. Here we show that antiferromagnetism in Ca$_{1-x}$La$_x$FeAs$_2$ is robust against and even enhanced by doping. Using $^{75}$As-nuclear magnetic resonance and nuclear quadrupole resonance techniques, we find that an antiferromagnetic order occurs below the N\'eel temperature $T_{\rm N}$ = 62 K at a high doping concentration ($x$ = 0.15) where superconductivity sets in at the transition temperature $T_{\rm c}$ = 35 K.  
Unexpectedly, $T_{\rm N}$ is enhanced with increasing doping, rising up to $T_{\rm N}$ = 70 K at $x$ = 0.24. The obtained phase diagram of this new system enriches the physics of iron-based high-$T_{\rm c}$ superconductors.

\end{abstract}

\pacs{}

\maketitle



The relationship between magnetism and superconductivity is an important issue in strongly correlated electron systems. In copper oxides, carrier doping into Mott insulators suppresses the magnetic order and induces high temperature superconductivity \cite{PALee}. In iron pnictides or selenides, superconductivity also emerges after the magnetic state is destroyed. For example, in $Ln$FeAsO (1111 system, $Ln$ = lanthanoid), substitution a few percent of F for O suppresses the antiferromagnetic order and induces superconductivity with the transition temperature $T_{\rm c}$ up to 55 K  \cite{Kamihara,Ren1,Ren2,Xu,Ren3,GFChen,Bos}. In BaFe$_2$As$_2$ (122 system), substitution of K ($TM$ = transition metal) for Ba (Fe) introduces hole (electron) to the system \cite{Rotter,Wang,LJLi,Canfield}. In either cases, superconductivity appears at low doping level less than 10\% \cite{Rotter,Wang,LJLi,Canfield}. In some of these compounds, magnetism can coexist microscopically with superconductivity \cite{Li,Zhou}. The extensive studies have shown that the quantum spin fluctuations associated with the magnetic order are important for superconductivity \cite{Zhou,Imai,Oka,Fukazawa}.

However, this picture  has been challenged by recent  materials which suggested that the phase diagram of the iron-pnictides can actually be much richer and the physics may be more diverse. With the high-pressure technique, it has been reported that one can dope carriers beyond $x =$ 0.5 \cite{Hosono,Yang}.  In LaFeAsO$_{1-x}$F$_x$, although no magnetism or even  magnetic  fluctuations was found in the high-doping region,  $T_{\rm c}$ forms another dome centered at $x \sim$ 0.5 \cite{Yang}. In  hydride-doped LaFeAsO$_{1-x}$H$_x$, $T_{\rm c}$ also forms another dome \cite{Hosono}, but  antiferromagnetism emerges after  superconductivity disappears  for $x > 0.5$ \cite{Fujiwara1,Hiraishi}. 
The magnetism in the highly hydride-doped system is intriguing,  but the origin of this antiferromagnetism and its relation to superconductivity are still unclear \cite{Fujiwara2,Hiraishi}. In particular, requirement of the unconventional method (a high-pressure synthesis technique) to obtain LaFeAsO$_{1-x}$H$_x$ samples prevents this interesting phenomenon far from understanding. A more easily-available material example is highly desired.

Recently, a new system, Ca$_{1-x}$La$_x$FeAs$_2$ (112-type), was discovered \cite{Katayama,Yakita,Kudo1,Kudo2} by the conventional synthesis technique. 
It crystallizes in a monoclinic structure (space group $P$2$_1$) and consists of the stacking of the Fe$_2$As$_2$ and the (Ca,La)$_2$As$_2$ layers along the $c$-axis. The pure compound CaFeAs$_2$ cannot be obtained, but replacing Ca with La ($x\geq$ 0.15) can  stabilize the 112 phase.  The highest $T_{\rm c}$ = 35 K was found for $x$ = 0.15 \cite{Katayama,Kudo1,Kudo2}. Since one La atom introduces one electron to the Fe$_2$As$_2$ layer, the doping level is high compared to the ambient-pressure-synthesized 1111 or 122 systems. With further doping beyond $x$ = 0.15, $T_{\rm c}$ decreases  and disappears at $x$ = 0.25 \cite{Kudo1}. The band structure calculation showed that the overall Fermi surfaces are similar to those of LaFeAsO$_{1-x}$F$_x$ \cite{Katayama}.  
Interestingly, it has been confirmed that both the As-Fe-As bond angle and the lattice parameter do not change from $x$ = 0.15 to 0.25 \cite{Kudo1,Kudo2}.

In this Rapid Communication, we report a finding of a magnetic order with the N\'eel temperature $T_{\rm N}$ = 62 K for Ca$_{0.85}$La$_{0.15}$FeAs$_2$ with $T_{\rm c}$ = 35 K by  $^{75}$As-nuclear magnetic resonance (NMR) / nuclear quadrupole resonance (NQR) techniques. Furthermore, we find unexpectedly that $T_{\rm N}$ increases with increasing doping, rising up to $T_{\rm N}$ = 70 K for $x$ = 0.24.  Our system provides a new opportunity to study the relationship between high-$T_{\rm c}$ superconductivity and doping-enhanced magnetism.

 Single crystals of Ca$_{1-x}$La$_x$FeAs$_2$ ($x$ = 0.15, 0.19, and 0.24) were prepared as reported elsewhere \cite{Katayama,Kudo1,Kudo2}. The La concentrations  were determined by energy-dispersive X-ray spectrometry measurement. ac susceptibility measurements using the NMR coil indicate $T_{\rm c}$ $\sim$ 35 and 34 K for $x$ = 0.15 and 0.19, respectively, but no $T_{\rm c}$ is observed for $x$ = 0.24. The magnetization $M$ was measured using the Quantum Design Magnetic Property Measurement System (MPMS).  In order to achieve a good signal to noise ratio, about 500 mg of small-sized (about 100 $\mu$m in diameter) single crystals were collected for NMR/NQR measurements \cite{Pulses}.  NMR/NQR spectra were taken by changing rf frequency and recording spin echo intensity step by step. The $^{75}$As-NMR $T_1$ was measured at the frequencies in the center transition ($m = 1/2\leftrightarrow-1/2$) peak \cite{Sup}.

 \begin{figure}
 \begin{center}
 \includegraphics[width=8.5cm]{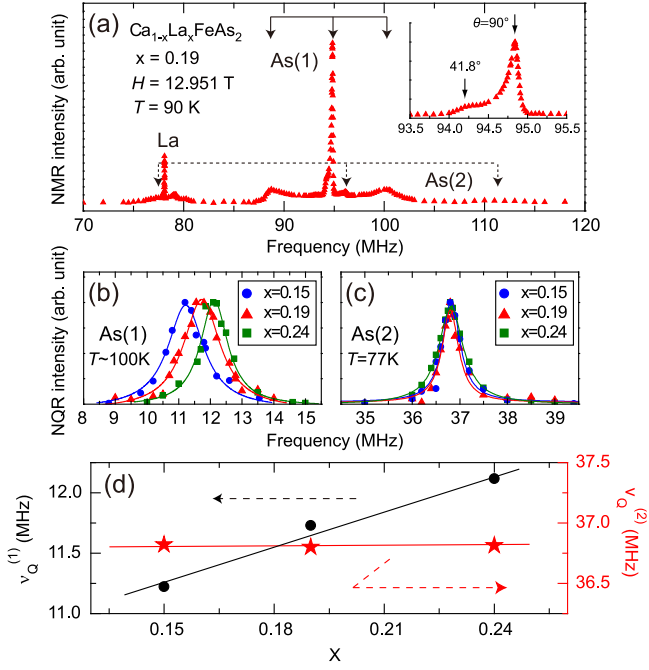}
 \end{center}
 \caption{(Color online) (a) NMR spectrum for $x$ = 0.19.  Inset shows the center peak for As(1).  (b) -- (c) $x$ dependence of the $^{75}$As-NQR spectra for As(1)  and As(2).  The solid lines are the results of Lorentzian fitting to obtain the peak position, $\nu_{\rm Q}$$^{\rm As(1,2)}$. (d) $x$ dependence of $\nu_{\rm Q}$$^{\rm As(1,2)}$. Solid lines are an eye guide.}
 \label{f1}
 \end{figure}

As a typical example, Fig. 1(a) shows the $^{75}$As-NMR ($I$ = 3/2) spectrum for $x$ = 0.19.  The nuclear spin Hamiltonian is expressed as a sum of the Zeeman and nuclear quadrupole interaction terms, $\mathcal{H}$ = $\mathcal{H}_{\rm z} + \mathcal{H}_{\rm Q}$ = $-^{75}\gamma\hbar\vec{I}\cdot\vec{H}_{0} (1+K) + (h \nu_{\rm Q}/6)[3{I_z}^2-I(I+1)+\eta({I_x}^2-{I_y}^2)]$, where  $^{75}\gamma$ = 7.292 MHz T$^{-1}$, $h$ is Planck's constant, $H_0$ is the external magnetic field, $K$ is the Knight shift, and $I$ is the nuclear spin. NQR frequency $\nu_{\rm Q}$ and asymmetry parameter $\eta$ are defined as $\nu_{\rm Q}$  = $\frac{3eQV_{zz}}{2I(2I-1)h}$, $\eta$ $=$ $\frac{V_{xx} - V_{yy}}{V_{zz}}$, with $Q$ and $V_{\alpha \beta}$ being the nuclear quadrupole moment and the electric field gradient (EFG) tensor at the As site, respectively \cite{Abragam}. As seen in Fig.1(a) inset, the $^{75}$As-NMR center transition consists of two peaks which correspond to crystallites with $\theta$ = 41.8$^\circ$ and 90$^\circ$, where $\theta$ is the angle between $H_0$ and the principal axis of the EFG along the $c$-axis \cite{Abragam}. Theoretically, the peak intensity of  $\theta$ = 41.8$^\circ$ is higher than that of  $\theta$ = 90$^\circ$ for a complete powder pattern \cite{Abragam}. However, we found the opposite, which indicates that about 50 \% of the tiny crystals are aligned to $H_0$, with $H_0\parallel$$ab$-plane. This situation allowed us to obtain $T_1$ with $H_0\parallel$$ab$-plane with great accuracy \cite{Sup}. As shown by the solid [As(1) site] and dotted [As(2) site] arrows, we obtained two sets of $^{75}$As-NMR spectra, since Ca$_{1-x}$La$_x$FeAs$_2$ has two inequivalent As sites, one in the Fe$_2$As$_2$ layer and the other in the (Ca,La)$_2$As$_2$ layer. Each set has three transitions from $I_{\rm z}$ = (2$m$$+$1)/2 to (2$m$$-$1)/2 where $m$ = $-$1, 0, 1.  From Fig. 1(a), we estimated $\nu_{\rm Q}^{\rm As(1)}\sim$ 11.5 MHz, $\nu_{\rm Q}^{\rm As(2)}\sim$ 35 MHz, and $\eta\sim$ 0 for both sites.

 \begin{figure}
\begin{center}
\includegraphics[width=7cm]{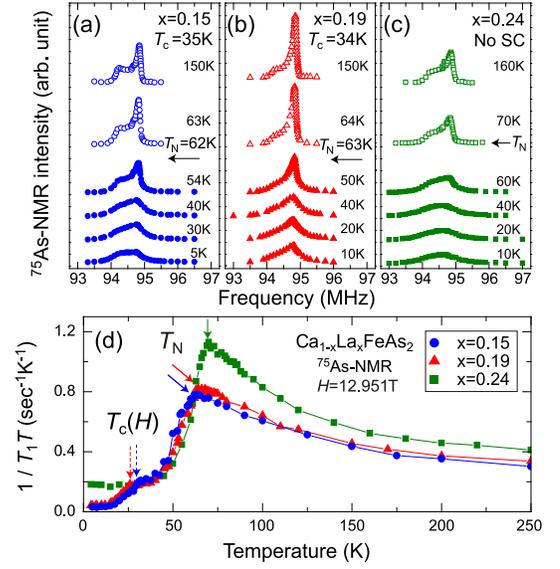}
\end{center}
\caption{(Color online) Temperature dependence of the $^{75}$As-NMR center peak for As(1) for $x$ = 0.15 (a), 0.19 (b), and 0.24 (c), respectively.  (d) Temperature dependence of 1/$T_1T$. Solid and dotted arrows indicate $T_{\rm N}$ and $T_{\rm c}(H)$. } 
\label{f2}
\end{figure}
 
 \begin{figure}
 \begin{center}
 \includegraphics[width=7cm]{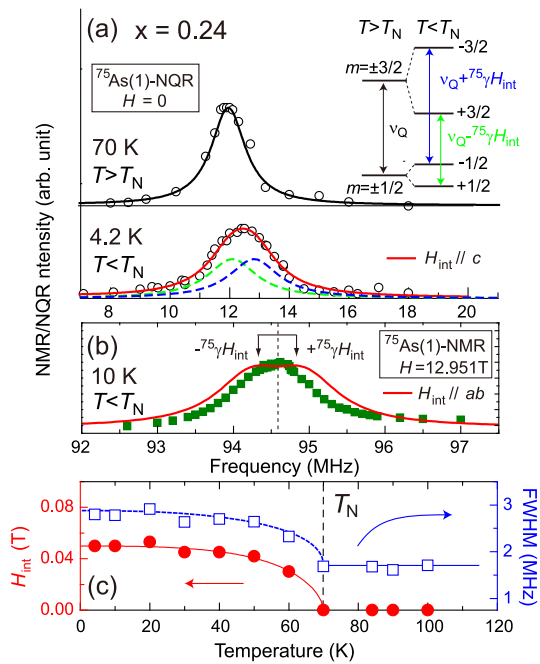}
 \end{center}
 \caption{(Color online)  (a) The $^{75}$As-NQR spectrum for As(1) above and below $T_{\rm N}$ for $x$ = 0.24. The dotted and solid curves are results of fitting (see text). Inset shows schematic view of the NQR-transition levels. (b) The $^{75}$As-NMR spectrum for As(1) below $T_{\rm N}$ for $x$ = 0.24. The dotted line indicates the $\theta = 90^\circ$ peak position ($f_\perp$). The solid curve is a sum of the two Lorentzian curves centered at  $f_\perp\pm ^{75}\gamma H_{\rm int}$.   (c) Temperature dependences of the FWHM of the $^{75}$As(1)-NQR spectrum and $H_{\rm int}$ for $x$ = 0.24. The dotted curve and solid horizontal line on FWHM($T$) are an eye guide. The solid line on $H_{\rm int}(T)$ is the mean-field fit. Dashed vertical line indicates $T_{\rm N}$. } 
 \end{figure}

 Figure 1(b) and 1(c) show the NQR spectra.  As summarized in Fig. 1(d),  $\nu_{\rm Q}^{\rm As(1)}$ increases with increasing doping, but, $\nu_{\rm Q}^{\rm As(2)}$ remains about the same. Since the lattice parameters do not change with doping \cite{Kudo1}, the origin of doping dependence of $\nu_{\rm Q}$ should be the EFG generated by the doped electrons in the Fe$_2$As$_2$ layer. Hence, we conclude that As(1) originates from the Fe$_2$As$_2$ layer and As(2) from the (Ca,La)$_2$As$_2$ layer. The doping dependence of $\nu_{\rm Q}^{\rm As(1)}$ is in good agreement with that reported in LaFeAsO$_{1-x}$F$_x$ \cite{Oka}.

To investigate the evolution of the normal-state electronic property, we measured the temperature dependence of the $^{75}$As-NMR center peak and $T_1$ for As(1).  As seen in Fig. 2, the $^{75}$As-NMR spectra unexpectedly became broad at low temperatures for all doping levels. Concomitantly,  1/$T_1$ divided by temperature (1/$T_1T$) showed a peak  at $T$ =  62,  63, and 70 K for $x$ = 0.15, 0.19, and 0.24, respectively. This is consistent with the critical slowing down behavior across $T_{\rm N}$ as observed in BaFe$_2$As$_2$ \cite{Kitagawa} and LaFeAsO$_{0.97}$F$_{0.03}$ \cite{Oka}, suggesting that the origin of the broadening of the NMR spectra is due to the appearance of the internal magnetic field $H_{\rm int}$ at the As site.

  In order to further confirm the origin of the spectrum broadening, we measured the evolution of the NQR spectrum and compared it with NMR result. Figure 3(a) shows the $^{75}$As-NQR spectrum obtained at $T$ = 70 K and 4.2 K for $x$ = 0.24.  Figure 3(b) shows the $^{75}$As-NMR spectrum at $T$ = 10 K for $x$ = 0.24 chosen from Fig. 2(c). As was observed in the NMR, the $^{75}$As-NQR spectrum also became broad at low temperatures.  Figure 3(c) shows the temperature dependence of the full width at the half maximum (FWHM) of  the $^{75}$As-NQR spectrum. To obtain the $H_{\rm int}(T)$ and $T_{\rm N}$, we employed the following simulation. In a magnetically ordered state, the NQR Hamiltonian is expressed as $\mathcal{H}_{\rm AFM} = -^{75}\gamma\hbar\vec{I}\cdot\vec H_{\rm int} + \mathcal{H}_{\rm Q}$. As a first step, we assumed  $\vec H_{\rm int} = (0, 0, H_{\rm int})$ as was found in the underdoped iron pnictide \cite{Kitagawa}. As shown schematically in Fig. 3(a) inset, the energy levels of the nuclear spins change and the $^{75}$As-NQR spectrum splits into two peaks at $\nu^{\rm AFM}$ = $\nu_{\rm Q}$ $\pm$ $^{75}\gamma H_{\rm int}$ below $T_{\rm N}$.  The NQR spectrum at $T$ = 4.2 K can be reproduced by a sum of the two Lorentzian curves centered at $\nu^{\rm AFM}$.  This indicates that the $H_{\rm int}$ below $T_{\rm N}$ is spatially uniform. On the other hand, if we assume the $\vec H_{\rm int}$ is parallel to the $ab$-plane, it can also reproduce the NQR spectrum with $H_{\rm int}$ = 0.05 T \cite{Sup}. However, as seen in Fig. 3(b),  $\vec H_{\rm int}$ $\parallel$ $ab$-plane, which generates an in-plane internal field of $\pm$$H_{\rm int}$ at the As site, is inconsistent with the NMR spectrum. Rather, it seems consistent with $\vec H_{\rm int} \parallel c$-direction, which just shifts the peak position and makes the NMR spectrum width broad \cite{Li}.    And hence, $\vec H_{\rm int} = (0, 0, H_{\rm int})$ below $T_{\rm N}$ is the most plausible in this case. The obtained  $H_{\rm int}(T)$ is plotted in Fig. 3(c).  The solid curve is a fit to the mean field theory that gives a $T_{\rm N}$ = 70 K. This is exactly the temperature at which $1/T_1T$ shows a peak.  From these results, we conclude that the peak in $1/T_1T$ is due to the onset of an antiferromagnetic order.
  
    As seen in Fig. 2(d), the antiferromagnetic order occurs in the highly electron-doped region ($x\geq$ 0.15) in the present system, while the antiferromagnetic phase is completely suppressed in $Ln$FeAsO$_{1-x}$F$_x$ \cite{Kamihara,Ren1,Ren2,Xu,Ren3,GFChen,Bos} and Ba(Fe$_{1-x}$$TM$$_x$)$_2$As$_2$ \cite{Wang,LJLi,Canfield} at much lower doping concentration. Notably, $T_{\rm N}$ increases with increasing doping. The mechanism of the increase of $T_{\rm N}$ by doping is unclear at the present stage. One possibility is that the nesting between enlarged Fermi surfaces due to electron doping \cite{Liu,Xie} can act to enhance the antiferromagnetism \cite{Onari}.

      \begin{figure}
      \begin{center}
     \includegraphics[width=7.5cm]{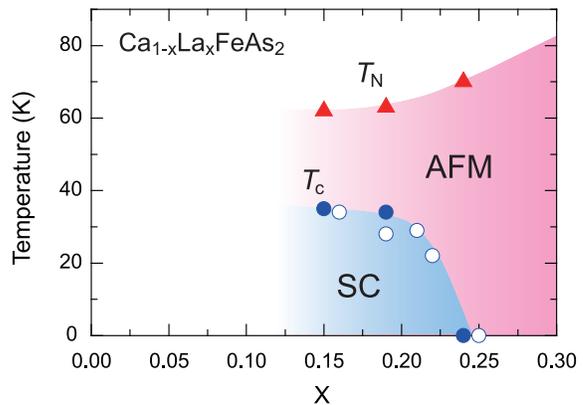}
      \end{center}
      \caption{(Color online) Phase diagram for Ca$_{1-x}$La$_x$FeAs$_2$.  AFM
         and SC denote the antiferromagnetically ordered metal and superconducting state. Open circles are from the magnetization results for another batch of samples \cite{Kudo1}. }
    
      \label{f4}
      \end{figure}

Next, we discuss possible magnetic structure below $T_{\rm N}$. According to the previous report \cite{Kitagawa}, the $\vec H_{\rm int}$  at the As site can be written as the sum of the contribution from the four nearest neighbor Fe sites as $\vec H_{\rm int} = \sum_{i=1}^4 \widetilde{B_i}\cdot \vec m_i$, where $\widetilde{B_i}$ is the hyperfine coupling tensor consisting of the components $B_{\alpha\beta}$ (\{$\alpha$, $\beta$\} = $a, b, c$) between As nucleus and $i$th Fe site, and $\vec m = (m_a, m_b, m_c)$ is the magnetic moment at the Fe site. First, we compare with the ordering vector $\vec q$ = ($\pi$, 0, 0) or ($\pi$, 0, $\pi$) found in the underdoped iron pnictides \cite{Kitagawa}. In this case, $\vec m$ produces $\vec H_{\rm int}$ = 4$B_{ac} (m_c, 0, m_a)$. Using $B_{ac}$ = 0.43 T/$\mu_{\rm B}$ \cite{Kitagawa}, we obtain $m_a$ $\sim$ 0.03 $\mu_{\rm B}$ which yield $H_{\rm int}$ = 0.05 T. However, this value would be one order magnitude smaller than that observed in LaFeAsO ($m_{\rm Fe}$ = 0.36 $\mu_{\rm B}$, $T_{\rm N}$ = 137 K) \cite{Cruz} and  BaFe$_2$As$_2$ ($m_{\rm Fe}$ = 0.87 $\mu_{\rm B}$, $T_{\rm N}$ = 143 K) \cite{Huang}. Below we consider another possibilities. The first one is $\vec q = (\pi, \pi, 0)$ or $(\pi, \pi, \pi)$. In this case, $\vec H_{\rm int}$ is perpendicular to the $c$ direction \cite{Kitagawa} and can be ruled out. The second one is $\vec q = (0, \pi, 0)$ or $(0, \pi, \pi)$, which produces $\vec H_{\rm int}$ = 4$B_{bc} (0, m_c, m_b)$. This type of magnetic structure has recently been discovered in LaFeAsO$_{0.49}$H$_{0.51}$ with  $m_{a}$ = 1.21 $\mu_{\rm B}$ and $T_{\rm N}$ = 89 K \cite{Hiraishi}. Obviously, this does not produce $H_{\rm int}$ at the As site. The observed spatially distributed $H_{\rm int}$ \cite{Fujiwara2} may be because that this antiferromagnetic state is accompanied by a structural transition, which causes displacement of As and Fe atoms about  0.13 \AA~ in the $a$ direction \cite{Hiraishi}.  Our system is similar to this situation. The As atom in Ca$_{1-x}$La$_x$FeAs$_2$ also exhibits small displacement of 0.03 \AA~ in the $a$ direction originated from its monoclinic crystal structure \cite{Katayama}. Therefore, although further quantitative understanding is difficult, this may explain the present results. In fact, such order was predicted by the band calculation for CaFeAs$_2$ \cite{XWu}. We call for neutron diffraction measurements to determine the magnetic structure for Ca$_{1-x}$La$_x$FeAs$_2$.

  \begin{figure}
   \begin{center}
   \includegraphics[width=8cm]{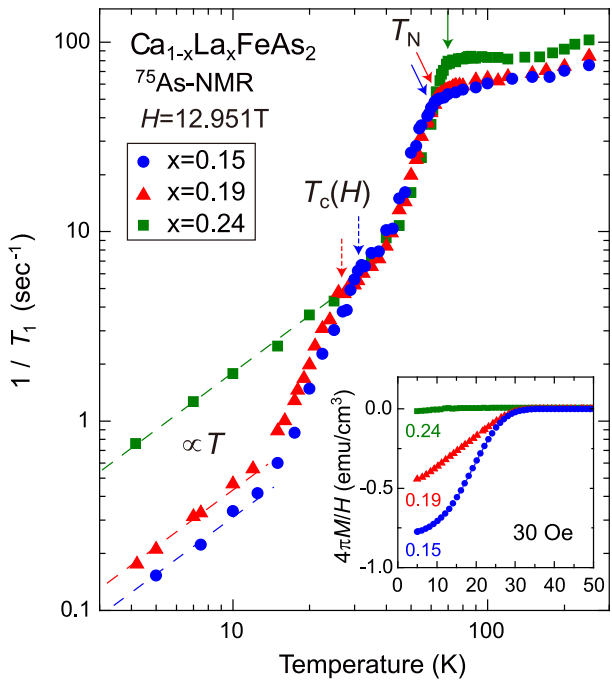}
   \end{center}
    \caption{(Color online) Temperature dependence of $1/T_1$. The solid and dotted arrows indicate $T_{\rm N}$ and $T_{\rm c}(H)$, respectively. The dashed lines indicate the relation $1/T_1T$ = const. Inset shows the temperature dependences of the magnetization. }
   \label{f5}
   \end{figure}

Figure 4 summarizes the phase diagram of  Ca$_{1-x}$La$_x$FeAs$_2$. We find that antiferromagnetism in Ca$_{1-x}$La$_x$FeAs$_2$ is robust against doping.  Furthermore,  $T_{\rm N}$ increases with increasing doping. This phase diagram is qualitatively similar to the heavily doped LaFeAsO$_{1-x}$H$_x$ \cite{Fujiwara1,Hiraishi}. However, there are two fundamental differences between them. First, the heavily doped LaFeAsO$_{1-x}$H$_x$ shows structural phase transition above $T_{\rm N}$ \cite{Hiraishi}, but it has not been observed in Ca$_{1-x}$La$_x$FeAs$_2$ \cite{Katayama}. Secondly, the situation regarding the relationship between antiferromagnetism and superconductivity is completely different. In heavily doped LaFeAsO$_{1-x}$H$_x$, the antiferromagnetic phase and superconductivity are segregated completely \cite{Fujiwara2}, but they coexist microscopically in Ca$_{1-x}$La$_x$FeAs$_2$ as elaborated below. 

Figure 5 shows the temperature dependence of $1/T_1$ \cite{Sup}. For all La concentration, $1/T_1$ decreases rapidly below $T_{\rm N}$ in accordance with the uniform $H_{\rm int}(T)$ (see Fig. 3). As shown in Fig. 5 inset, the bulk nature of superconductivity for $x$ = 0.15 and 0.19 is confirmed by the large superconducting shielding fractions and further assured by the distinct reduction of $1/T_1$ below $T_{\rm c}$. These results indicate that antiferromagnetism and superconductivity coexist microscopically for $x$ = 0.15 and 0.19.  Furthermore, $1/T_1$ shows  1/$T_1T$ = const. behavior well below $T_{\rm N}$ for $x$ = 0.24, suggesting that the heavily doped Ca$_{1-x}$La$_x$FeAs$_2$ is an antiferromagnetic metal. For $x$ = 0.15 and 0.19, $1/T_1$ decreases rapidly again below $T_{\rm c}(H)$ as found previously \cite{Imai,Oka,Zhou,Fukazawa}. It is worth noting that $1/T_1$ at low temperatures in the superconducting state becomes to be proportional to $T$. In the superconducting state coexisting with an antiferromagnetic order, the spin rotation is broken so that a spin-triplet component can be mixed \cite{Chubukov}. Our observation may be a reflection of such novel phenomenon, since the spin triplet state will have nodes in the gap function. More work in  this regard  is deserved.

In summary, we have presented the systematic NMR/NQR studies on the La-doped iron-pnictide Ca$_{1-x}$La$_x$FeAs$_2$. We find the doping-enhanced antiferromagnetism that microscopically coexist with superconductivity. Such feature has never been reported before among the iron pnictides and/or the high $T_{\rm c}$ cuprates. Since the lattice parameters in Ca$_{1-x}$La$_x$FeAs$_2$ do not change by doping, the unusual evolution of antiferromagnatism and superconductivity compared with other iron pnictides is originated purely from doped carriers. The present results provide further opportunity and new perspective for understanding the roll of doped carrier and the mechanism of high $T_{\rm c}$ superconductivity in the iron pnictides.

We thank S. Onari for discussion. This work was supported in part by research grants from MEXT (No. 22103004, 25400372, 25400374, and 26287082) and by CAS.



\begin{thebibliography}{9}
\bibitem{PALee}
P. A. Lee $et$ $al$., Rev. Mod. Phys. {\bf 78}, 17 (2006).
\bibitem{Kamihara}
Y. Kamihara $et$ $al$., J. Am. Chem. Soc. {\bf130}, 3296 (2008).
\bibitem{Ren1}
Z.-A. Ren $et$ $al$.,  EPL {\bf 82} 57002 (2008).
\bibitem{Ren2}
Z.-A. Ren $et$ $al$., Mater. Res. Innovations {\bf 12} 106 (2008).
\bibitem{Xu}
C. Wang $et$ $al$., EPL {\bf 83}, 67006 (2008). 


\bibitem{Ren3}
Z.-A. Ren $et$ $al$., Chin. Phys. Lett. {\bf 25}, 2215 (2008).

\bibitem{GFChen}
G. F. Chen $et$ $al$., Phys. Rev. Lett. {\bf 100}, 247002 (2008).

\bibitem{Bos}
J.-W. G. Bos $et$ $al$.,  Chem. Commun. 3634 (2008).


\bibitem{Rotter}
M. Rotter $et$ $al$., Angew. Chem. Int. Ed. {\bf 47}, 7947 (2008).

\bibitem{Wang}
X. F. Wang $et$ $al$., New J. Phys. {\bf 11}, 045003 (2009). 
 
 \bibitem{LJLi}
 L. J. Li $et$ $al$., New J. Phys.{\bf 11}, 025008 (2009).
 
 \bibitem{Canfield}
 P. C. Canfield $et$ $al$., Phys. Rev. B {\bf 80}, 060501(R) (2009). 
 
 \bibitem{Li}
 Z. Li $et$ $al$., Phys. Rev. B {\bf 86}, 180501(R) (2012).
 
 
 \bibitem{Zhou}
    R. Zhou $et$ $al$., Nat. Commun. {\bf 4}, 2265 (2013). 
 
 \bibitem{Imai}
 F. L. Ning $et$ $al$., Phys. Rev. Lett. {\bf 104}, 037001 (2010).
 
 
  \bibitem{Oka}
  T. Oka $et$ $al$., Phys. Rev. Lett. {\bf 108}, 047001 (2012).
  
\bibitem{Fukazawa}
M. Hirano $et$ $al$., J. Phys. Soc. Jpn. {\bf 81}, 054704 (2012).  

\bibitem{Hosono}
S. IImura $et$ $al$., Nature Commun. {\bf 3}, 943 (2012).

\bibitem{Yang}
J. Yang $et$ $al$., Chin. Phys. Lett. {\bf 32}, 100601 (2015). (also available at arXiv:1507.01750).
\bibitem{Fujiwara1}
N. Fujiwara $et$ $al$., Phys. Rev. Lett. {\bf 111}, 097002 (2013).    
   
\bibitem{Hiraishi}
M. Hiraishi $et$ $al$., Nat. Phys. {\bf 10}, 300 (2014).
 
  
\bibitem{Fujiwara2}
R. Sakurai $et$ $al$., Phys. Rev B {\bf 91}, 064509 (2015).

\bibitem{Katayama}
N. Katayama $et$ $al$., J. Phys. Soc. Jpn. {\bf 82}, 123702 (2013).


\bibitem{Kudo1}
K. Kudo $et$ $al$., J. Phys. Soc. Jpn. {\bf 83}, 025001 (2014).

\bibitem{Kudo2}
K. Kudo $et$ $al$.,  J. Phys. Soc. Jpn. {\bf 83}, 093705 (2014).

\bibitem{Yakita}
H. Yakita $et$ $al$.,  J. Am. Chem. Soc. {\bf 136}, 846 (2014). 

\bibitem{Pulses}
NMR/NQR measurements were carried out by using a phase-coherent spectrometer and at a fixed magnetic field of $H$ = 12.951 T. Typical pulse sequence 10~$\mu$s -- $\tau$ -- 20~$\mu$s with $\tau$ = 25 $\mu$s is used to obtain spin echo.

\bibitem{Sup}
See Supplemental Material for the recovery curves to obtain $T_1$ and the result of simulation with  $\vec H_{\rm int} \parallel ab$-plane. 

\bibitem{Abragam}
A. Abragam, $The Principles of Nuclear Magnetism$ (Oxford University Press, London, 1961).


\bibitem{Kitagawa}
K. Kitagawa $et$ $al$., J. Phys. Soc. Jpn. {\bf 77}, 114709 (2008). 



\bibitem{Liu}
X. Liu $et$ $al$., Chin. Phys. Lett. {\bf 30}, 127402 (2013).  

\bibitem{Xie}
M. Y. Li $et$ $al$., Phys. Rev. B {\bf 91}, 045112 (2015).

\bibitem{Onari}
S. Onari $et$ $al$., Phys. Rev. Lett. {\bf 112}, 187001 (2014).

\bibitem{Cruz}
Clarina de la Cruz $et$ $al$., Nature {\bf 453}, 899 (2008).

\bibitem{Huang}
Q. Huang $et$ $al$., Phys. Rev. Lett. {\bf 101}, 257003 (2008).

\bibitem{XWu}
X. X. Wu $et$ $al$., Phys. Rev. B {\bf 89}, 205102 (2014).
 

 
\bibitem{Chubukov}
A. Hinojosa $et$ $al$., Phys. Rev. Lett. {\bf 113}, 167001 (2014).

\end{thebibliography}
\end{document}